\documentclass[aps,prx,twocolumn,showpacs,amsmath,amssymb,superscriptaddress,longbibliography]{revtex4-2}

\usepackage{graphicx}

\usepackage{amssymb}
\usepackage{amsmath}
\usepackage{bm}
\usepackage{color}
\usepackage{enumitem}
\usepackage{multirow}

\newcommand{\lla}{\left\langle}
\newcommand{\rra}{\right\rangle}
\newcommand{\kappard}{\lambda}
\newcommand{\phimean}{\phi}
\newcommand{\nsq}{N_{sq}}

\graphicspath{{./Figures/}{./}}
\begin{document}
\title{Active turbulence in microswimmer suspensions — the role of active hydrodynamic stress and volume exclusion}
\author{Kai Qi}
\altaffiliation{Current address: CECAM Centre Europ{\'e}en de Calcul Atomique et Mol{\'e}culaire,
{\'E}cole Polytechnique F{\'e}d{\'e}rale de Lausanne (EPFL),  1015 Lausanne, Switzerland.}
\affiliation{Theoretical Physics of Living Matter, Institute of Biological Information Processing, and Institute for Advanced Simulation, Forschungszentrum J\"ulich, 
52425 J\"ulich, Germany}
\author{Elmar Westphal}
\affiliation{Peter Gr\"unberg Institute and J\"ulich Centre for Neutron Science, 
Forschungszentrum J\"ulich, D-52425 J\"ulich, Germany}
\author{Gerhard Gompper}
\email{g.gompper@fz-juelich.de}
\affiliation{Theoretical Physics of Living Matter, Institute of Biological Information Processing, and Institute for Advanced Simulation, Forschungszentrum J\"ulich, 
52425 J\"ulich, Germany}
\author{Roland G. Winkler}
\email{r.winkler@fz-juelich.de}
\affiliation{Theoretical Physics of Living Matter, Institute of Biological Information Processing, and Institute for Advanced Simulation, Forschungszentrum J\"ulich, 
52425 J\"ulich, Germany}
\begin{abstract}
\noindent{\bf  ABSTRACT} \\[5pt]
Microswimmers exhibit an intriguing, highly-dynamic collective motion with large-scale  swirling and streaming patterns, denoted as active turbulence — reminiscent of classical high-Reynolds-number hydrodynamic turbulence.  Various experimental, numerical, and theoretical approaches have been applied to elucidate similarities and differences to inertial  hydrodynamic and active turbulence. These studies reveal a wide spectrum of possible structural and dynamical behaviors of active mesoscale systems, not necessarily consistent with the predictions of the  Kolmogorov-Kraichnan theory of turbulence. We use squirmers embedded in a mesoscale fluid, modeled by  the multiparticle collision dynamics (MPC) approach, to explore the collective behavior of bacteria-type microswimmers. Our model includes the active hydrodynamic stress generated by propulsion, and a rotlet dipole characteristic for flagellated bacteria. We find emergent clusters, activity-induced phase separation, and swarming, depending on density, active stress, and the rotlet dipole strength. The analysis of the squirmer dynamics in the swarming phase yields Kolomogorov-Kraichnan-type hydrodynamic turbulence and  energy spectra for sufficiently high concentrations and strong rotlet dipoles. This emphasizes the paramount importance of the hydrodynamic flow field  for swarming and bacterial turbulence.

\end{abstract}
\maketitle 

\section*{Introduction} \label{sec:introduction}

Active matter  comprises a unique class of systems with intricate structural and dynamical features, facilitated by  their elementary agents consuming internal energy, or energy from the environment, to maintain an out-of-equilibrium state.  The interplay between the autonomous locomotion of the agents and their interactions leads to large-scale self-organized  collective motion manifested in such  diverse biological systems as flocks of birds \cite{vics:12,tone:95,cava:14,chat:20}, school of fish \cite{ward:08,shae:20}, bacterial colonies \cite{henr:72,soko:07,berg:04,cope:09,darn:10,kear:10,wens:12,marc:13,elge:15,beer:20}, epithelial cell monolayers \cite{haki:17,aler:20,tan:20}, and the cell cytoskeleton \cite{juel:07,need:17,doos:18,opat:19}, as well as  synthetic systems like robots \cite{cham:17,rube:14}, self-assembled magnetic spinners \cite{koko:17}, and phoretic colloids \cite{cohe:14,bour:20,gomp:20}.  

Swarming bacteria \cite{zhan:09.1,darn:10,domb:04,wolg:08,dunk:13,bepp:17,soko:12,wens:12,beer:20}, tissue cells \cite{pouj:07,doos:15,tan:20,lin:21}, and filament/motor-protein mixtures --- active nematics --- \cite{giom:15,brat:15,doos:18,opat:19,aler:20} exhibit a particular type of collective, chaotic  motion often denoted as {\em active turbulence} or {\em mesoscale turbulence},  with large-scale spatially and temporally random flow patterns. At first glance, the flow patterns are reminiscent  of those observed in classical high-Reynolds-number  hydrodynamic turbulence \cite{kolm:91,krai:80,fris:95}, despite active turbulence  occurring at exceedingly small Reynolds numbers. The similarity prompted intensive studies of the collective motion of active matter systems to unravel the underlying physical mechanisms due to its prototypical character for  nonlinear and nonequilibrium dynamical systems, which is considered  as a major challenge for current theoretical physics \cite{brat:15}.     

\setlength{\tabcolsep}{0.5em}
\begin{table*}[t] 
\caption{{\bf Various aspects of experimentally, theoretically, and by simulations studied systems exhibiting features of active turbulence.}
The articles (Ref.) discuss microscale systems exhibiting a power-law energy spectrum $E(k) \sim k^{-\kappa}$  for $k_{m} < k <  k_c$ and $E(k) \sim k^{\hat \kappa}$ for $k < k_{\rm m}$, with $k_{\rm m}$ and $k_c$ the wavenumber of the maximum in the energy spectrum and that of the microswimmer characteristic length, respectively. Note that in active nematic theory, $k_m = 2\pi/l_a$ \cite{giom:15,aler:20}. Cells comprise canine kidney, endothelial, myoblast, and fibroblast cells. Abbreviations: act. nem.: active nematics, SPR: self-propelled rod,   LB: Lattice Boltzmann, MPC: multiparticle collision dynamics,  HI: hydrodynamic interactions, $\phi$: packing fraction.  Symbols: ``$\checkmark$'' aspect is present, ``$-$'' aspect is absent, ``$/$'' aspect has not been analyzed/considered .\\}
\begin{tabular}{l | c | c c c  c c c c c  c}
\multirow{2} {*}{Technique} & \multirow{2} {*}{System/Approach} & \multirow{2} {*}{shape} & \multirow{2} {*}{HI} & excl. & active  & rotlet  &  Gaussian  & $\kappa, \ k^{-\kappa}$ & $\hat \kappa, \ k^{\hat \kappa}$    &  \multirow{2} {*}{Ref.} \\
 &  &  &  &  volume & stress &  dipole &   vel. distr. & (large $k$)& (small $k)$  &  \\ \hline \hline
\multirow{3} {*}{Experiment}  &  {\em B. subtilis} & elong. & $\checkmark$ & $\checkmark$ & $\checkmark$ &  $\checkmark$ & $\checkmark$ &$8/3$ & 5/3 &\cite{wens:12} \\
  &  {\em E. coli} & elong. & $\checkmark$ & $\checkmark$ & $\checkmark$ &  $\checkmark$ &  $/$ & $4/3$ & $3/5$ &\cite{bepp:17}  \\
  &  cells  & elong. & $\checkmark$ & $\checkmark$ & $\checkmark$ &  $-$ &  $-$ & $\gtrsim 13/3$ & $-$ &\cite{lin:21} 
     \\ \hline
\multirow{3} {*}{Theory} & field, isotrop.  & $/$ & $\checkmark$ & $-$ & $-$ &  $-$ &  $/$ & $8/3$ & 5/3 &\cite{wens:12} \\ 
 & \! \! act. nem. (def. free) \!\! &  $/$    &$\checkmark$ & $-$ & $-$ &  $-$ & $/$ & $12/3$ & $-1$ & \cite{aler:20} \\ 
 & act. nem.   &  $/$    &$\checkmark$ & $-$ & $-$ &  $-$ & $\checkmark$ & $12/3$ & $-1$ & \cite{giom:15} \\ \hline
\multirow{5} {*}{Simulations} & SPR  & rod & $-$ & $\checkmark$ & $-$ &  $-$ & $\checkmark$ & $\gtrsim 5/3$ & $/$ & \cite{wens:12.1} \\
& Vicsek-type  &  point part.&$-$ & $-$ & $ (\checkmark) $ &  $-$ & $/$ & $8/3$ &  $5/3$ & \cite{gros:14} \\ 
& LB & point part.& $\checkmark$ & $-$ & $\checkmark$ &  $-$ & $/$ & $11/3$ &  $/$ & \cite{bard:19} \\
& MPC: $\phi=0.60$   & spheroid & $\checkmark$ & $\checkmark$ & $\checkmark$ &  $\checkmark$ & $-$ &  $2$ & $5/3$  & this work\\
& MPC: $\phi=0.68$ & spheroid & $\checkmark$ & $\checkmark$ & $\checkmark$ &  $\checkmark$ & $\checkmark$ &  $5/3$ & $1$  & this work \\
\end{tabular}
\label{tab:energy_spectrum} 
\end{table*} 

Fundamental insight into  hydrodynamic  turbulence is achieved via velocity correlation functions \cite{batc:59}. In particular, Kolmogorov predicted the universal power-law dependence  for the energy spectrum $E  \sim  k^{-\kappa}$ on the wavenumber $k=|\bm k|$, with $\kappa = 5/3$ \cite{kolm:91,batc:59}. In fact, this relation  applies for two- (2D) and three-dimensional (3D) systems \cite{krai:80}. 
Numerous studies on active systems reveal a wide spectrum of possible turbulent characteristics  dependent on their constituents and the detailed (microscopic) interaction mechanisms, reflected in a wide range of exponents deviating from the Kolmogorov value, see Tab.~\ref{tab:energy_spectrum}.   Experiments on {\em B. subtilis} and {\em E. coli} bacteria \cite{bepp:17,wens:12} yield exponents significantly above and below the Kolmogorov value. Computer simulations employing various models have been performed and the energy spectrum has been calculated. Nonhydrodynamic particle-based simulations of an extension of the Vicsek model \cite{vics:95},  accounting for short-range parallel  and large-range antiparallel alignment, yield the same  exponent \cite{gros:14} as in experiments on {\em E. coli} \cite{wens:12}. Simulations of self-propelled rodlike particles  give a value close to the  Kolmogorov value \cite{wens:12.1,wens:12}.  Lattice Boltzmann simulations of microswimmers represented by extended force dipoles  (point particles) produce seemingly turbulent behavior  for sufficiently large swimmer densities \cite{bard:19} (see Table~\ref{tab:energy_spectrum}). For active nematics,  the route to chaotic behavior has been studied experimentally and theoretically \cite{doos:17,opat:19}. Their dynamics is characterized by an intrinsic length scale $l_a$, where $l_a$ is determined by the balance between the active and nematic elastic stress \cite{giom:15,aler:20}, and the creation and annihilation of topological defects.  In addition, various theoretical studies have been performed  with \cite{giom:15} and without \cite{aler:20} defects, where both yield similar energy spectra with distinct power-law exponents for length scales larger and smaller than $l_a$  (Tab.~\ref{tab:energy_spectrum}).  In contrast, we expect hydrodynamic interactions to dominate the chaotic and turbulent behavior in bacterial suspensions. Hence, it is a priori not evident that both types of chaotic dynamics exhibit the same kind of turbulent behavior, taken into account the disparity  in the exponents $\kappa$ and $\hat \kappa$.    

There are two particular systems of mesoscopic active particles, namely spinners --- short rodlike self-organized colloidal structures rotated by an external magnetic field \cite{koko:15} ---  and Marangoni surfers \cite{bour:20}, where turbulent dynamics consistent with Kolmogorov scaling has been observed. Their Reynolds numbers  $Re \sim {\cal{O}}(10)$ are much smaller than that of classical inertial turbulence,  but are much larger than  those of microswimmer systems, where  $Re \ll 1$. 

As a major difference to hydrodynamic turbulence, various experimental and simulation studies of active turbulence suggest the presence of a characteristic upper length scale for the  vortex size, only below which the energy spectrum decreases in a power-law  manner with increasing wavenumber  $k$ \cite{soko:12,domb:04,dunk:13}. This  scale is typically on the order of ten microswimmer lengths. Theoretical studies based on a continuum approach \cite{wens:12,dunk:13,brat:15,rein:18}, where the velocity field is described by the incompressible Toner-Tu equation \cite{tone:98,rama:10}  combined with a Swift-Hohenberg term \cite{swif:77} for pattern formation, support this observation.   However, in contrast to high-Reynolds number hydrodynamic turbulence, the internal stress due to self-propulsion and polar alignment interactions of the active agents is important, which, combined with the fluid dynamics,  determines the vortex size \cite{rein:18}.  


The diversity of obtained energy spectra and characteristic power laws (Tab.~\ref{tab:energy_spectrum}) indicates a strong dependence of the collective behavior on the detailed microswimmer interactions. Yet, it is not clear to which extent and under what circumstances hydrodynamic interactions are important. 

In this article, we perform extensive coarse-grained  mesoscale hydrodynamic simulations by  employing the multiparticle collision dynamics (MPC) approach for fluids \cite{kapr:08,gomp:09,thee:18} to elucidate the collective, turbulent motion of microswimmers  in monolayer films. The microswimmers are described in a coarse-grained manner applying the squirmer model \cite{ishi:06,pago:13,thee:16.1,zoet:18,thee:18}. Particular attention is paid to the influence of the microswimmers' hydrodynamic flow field on their collective behavior, i.e, the active stress  and the rotlet dipole resulting from the rotating of a flagella (bundles) and the counterrotating cell body in flagellated bacteria \cite{dres:11,hu:15.1,lope:14,ishi:20}. In general,  hydrodynamics plays a decisive role in the collective behavior of microswimmers. While dry spherical active Brownian particles (ABPs) exhibit motility-induced phases separation (MIPS) \cite{cate:15,elge:15,bech:16,bial:12,redn:13,wyso:14,digr:18}, microswimmers in the presence of hydrodynamics show cluster formation \cite{thee:18}, but no phase separation \cite{thee:18,mata:14}.  However, anisotropic, spheroidal squirmers exhibit enhanced clustering compared  to similar  ABP systems due to hydrodynamic attraction \cite{thee:18}. Hence, it is important to unravel the effect of shape, active stress, and of a rotlet dipole in dense microswimmer systems on their emergent collective properties, since bacteria in films exhibit swarming --- a rapid, coherent group migration over surfaces in dense populations,  with  large-scale swirling and streaming patterns \cite{cope:09,darn:10,beer:19,beer:20} ---  rather than clustering and phase separation \cite{zhan:09.1,darn:10,domb:04,wolg:08,dunk:13,bepp:17,soko:12,wens:12,beer:20}.  


By systematically varying the squirmer density, the active stress, and the rotlet dipole strength, our simulations provide insight into their influence on the collective dynamics of microswimmers. The combination of active stresses and a non-zero rotlet dipole suppresses phase separation and promotes swarming motility.

The analysis of the  swarming phases  reveals turbulent-like motion,  where the energy spectrum displays power-law decays below the characteristic length scale discussed above, however, with an exponent depending on the squirmer concentration. Remarkably, we find the value $\kappa =5/3$ for our largest density, strong active stress, and a non-zero rotlet dipole, consistent with the Kolmogorov prediction.

\begin{figure}[t]
\centering
  \includegraphics[width= \columnwidth]{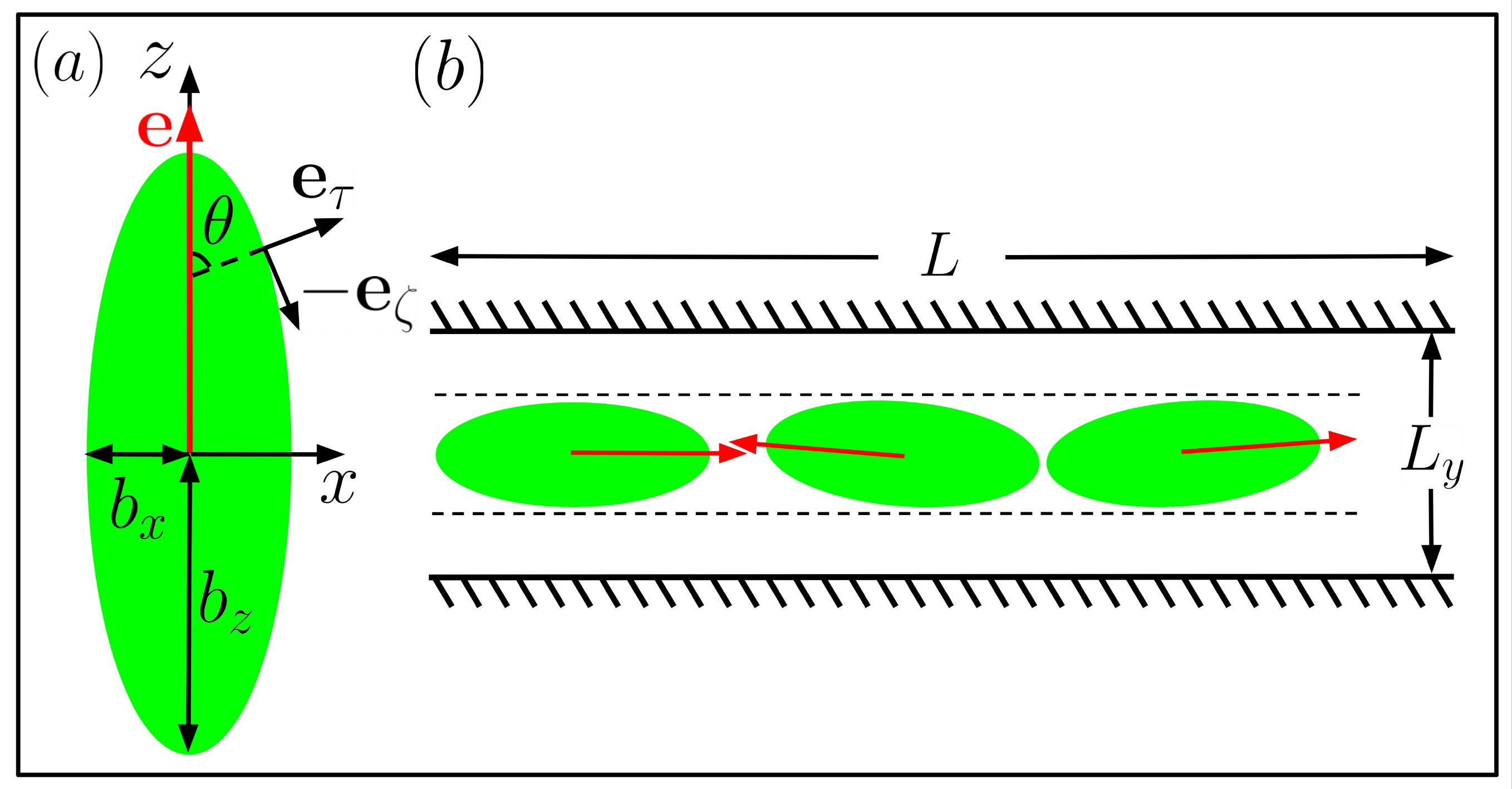}
  \caption{{\bf Illustration of the simulation setup.} (a) Sketch of a spheroidal squirmer, which is propelled in the direction ${\bm e}$  (red arrow) along the $z$-axis of the body-fixed reference frame. The spheroid's semi-major- and -minor axis are  $b_z$ and $b_x$, respectively, and  ${\bm e}_{\tau}$ and ${\bm e}_{\zeta}$ indicate the local normal and tangential unit vectors. (b) Multiple spheroidal squirmers in a narrow square-shaped slit of width $L_y = 4 b_x$ and lateral  extension $L$. A strong repulsive wall potential, as indicated by the dashed lines, implies quasi-2D confinement in the channel center. }
  \label{fig:sketch}
\end{figure}

\section*{Results}

\subsection*{System setup}

In our simulations,  $N_{sq}$ prolate spheroidal squirmers with the semi-major, $b_z$,  and -minor, $b_x$, axis  are confined in a three-dimensional narrow slit between two parallel walls and periodic boundary conditions  along the $x$ and $z$ direction (Fig.~\ref{fig:sketch}). The  prescribed squirmer surface velocity yields swimming with the velocity $v_0$, an active stress of strength $\beta$, and a rotlet dipole of strength $\lambda$ (Sec.~\ref{sec:method}). The embedding fluid is modeled via the multiparticle collision dynamics (MPC) method \cite{kapr:08,gomp:09}, applying the stochastic-rotation variant with angular momentum  conservation (MPC-SRD+a)  \cite{thee:16,nogu:08}.

\subsection*{Structural properties } \label{sec:structure}

\begin{figure}[t]
\centering
  \includegraphics[width= \columnwidth]{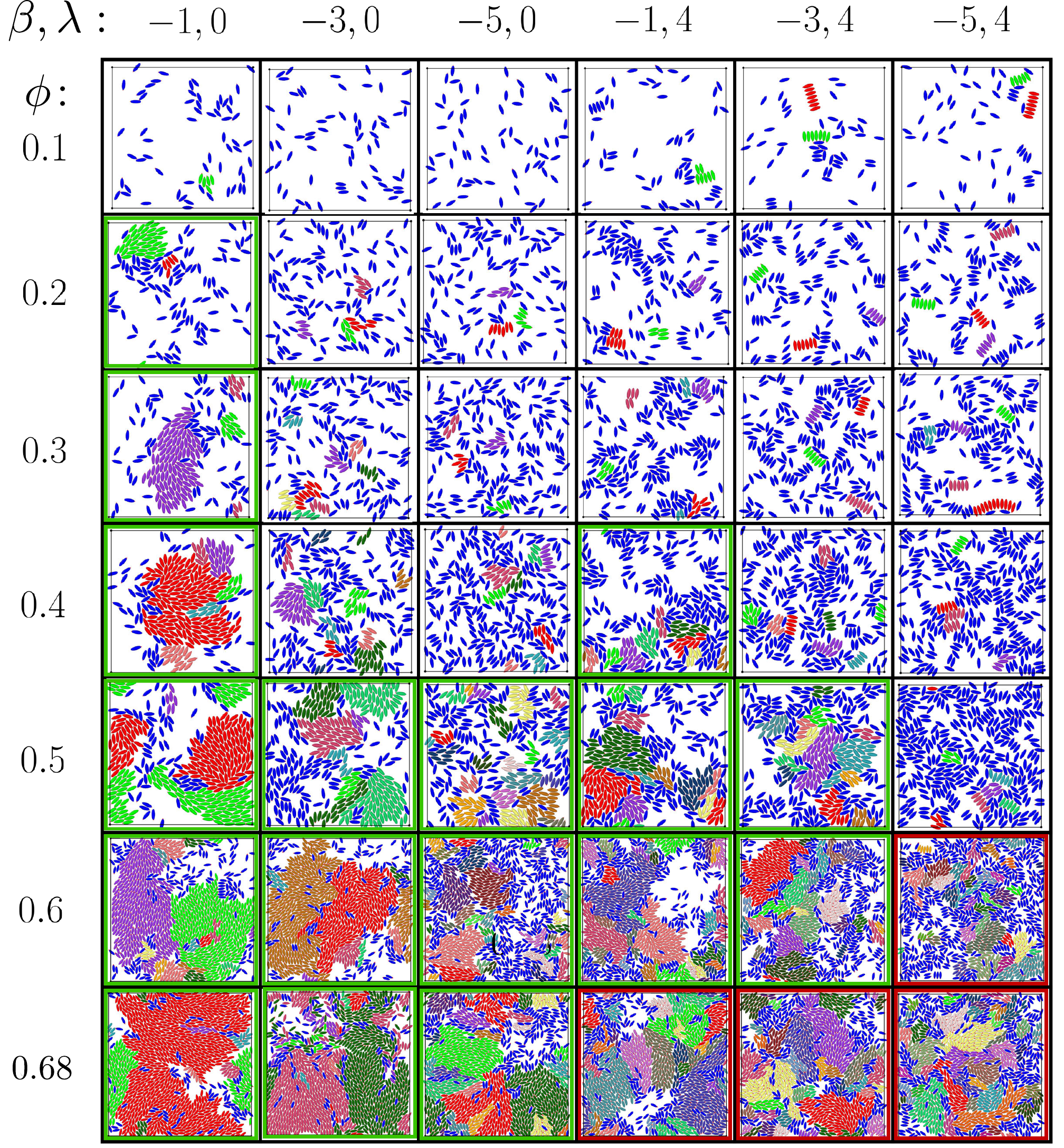}
  \caption{{\bf Snapshots of emergent structures.} Structures of squirmers  for various packing fractions, $\phi$, active stresses, $\beta$, and rotlet dipole strengths, $\lambda$.  The box sizes are $L=160a$  for $\phi \le 0.5$ and $L=230a$  for $\phi>0.5$. Small clusters with squirmer numbers $m \le 4$ are colored in blue, various other (random) colors are used for clusters with  $m > 4$. The snapshots with  green frames correspond to (large) clusters and A-MIPS, where clusters are systems-spanning at higher packing fractions  (see movies M1, M2). The snapshots with  red frames correspond to swarming systems (see movie M3). The other systems show individually squirmers and (few) small clusters (see movie M4). 
   }
  \label{fig:phase_diagram}
\end{figure}

\begin{figure}[t]
\centering
  \includegraphics[width= \columnwidth]{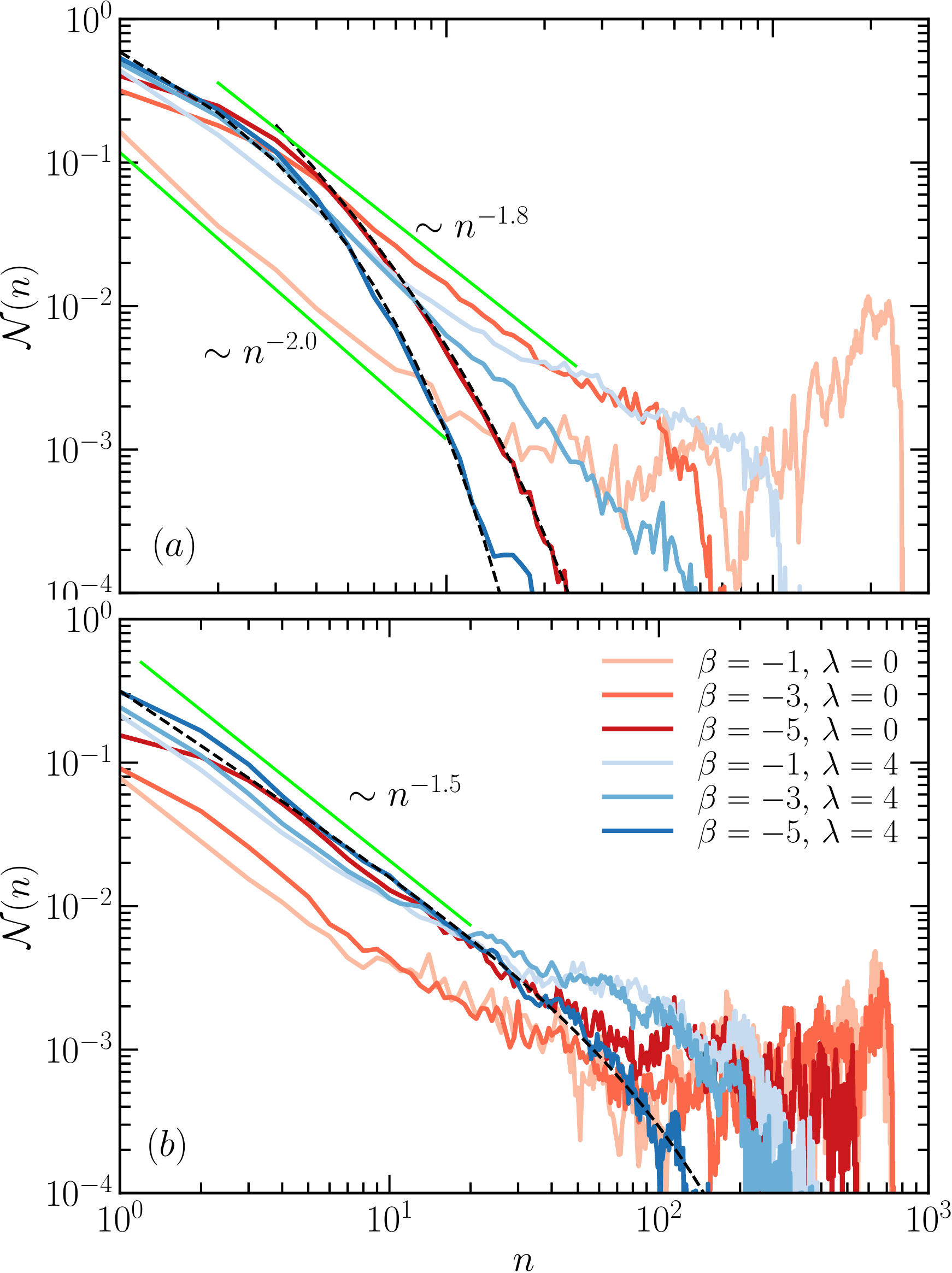} 
  \caption{{\bf Cluster-sizer distribution function.} Cluster-size distribution function $\mathcal{N}(n)$ \eqref{cl_size_distr_eq} for the average packing fractions (a) $\phi = 0.4$ and (b) $\phi = 0.6$.  The curves  present  results for $\beta = -1$, $-3$, and $-5$ (bright to dark) and $\lambda = 0 $ (red) and $\lambda = 4$ (blue), respectively.   The dashed lines are fits of the function $\mathcal{N}(x)$ of Eq.~\eqref{eq:cluster_size_fit} with the parameters of Tab.~\ref{tab:cluster_size}. The green solid lines indicate power laws with the respective exponents.  }
  \label{fig:cluster_size_distr}
\end{figure}

The  simulation snapshots of Fig.~\ref{fig:phase_diagram} illustrate emergent structures  for the various considered packing fractions, active  stresses, and rotlet dipole strengths. Distinct motility patterns can be identified: (i) Motility-induced phase separation (A-MIPS)  for $|\beta|\ge 1, \lambda=0, \phi \gtrsim 0.3$. Since here the shape of the spheroids implies squirmer alignment and the formation of polar motile clusters, we use the notation A-MIPS, to distinguish it from the case of isotropic, non-aligning particles, which form immobile clusters (MIPS) \cite{bial:12,elge:15,bech:16}. (ii) Swarming motility  for $|\beta| > 1, \lambda=4,  \phi \gtrsim 0.5$, and (iii) gas of small clusters  for $\phi \lesssim 0.3$. The clusters emerging by A-MIPS  increase with increasing packing fraction and are system-spanning for $\phimean \gtrsim 0.5$, consistent with our previous studies \cite{thee:18};  they are denoted as local and global clusters in Ref.~\cite{beer:20}.  The clusters are rather dynamic and exhibit translational and rotational motion.  
In the  dense swarming phase, clusters of squirmers migrate collectively, thereby  forming dynamic swirling and streaming patterns \cite{cope:09,kear:10,beer:19,beer:20}.  A quantitative criterion for the  classification into  A-MIPS and swarming motility  will be provided in terms of the cluster-size distribution function (Fig.~\ref{fig:cluster_size_distr}).  Some of the  small clusters for $\phi \lesssim 0.3$  exhibit  cooperative motion, where a few squirmers move together for some time. In general, the rotlet dipole enhances cluster formation, and squirmers align side by side, which is clearly visible for  $\phi \lesssim 0.4$. The precise mechanism for this cooperative motion is unexplored, but could depend on squirmer wall interactions. In contrast,  for larger packing fractions the rotlet dipole suppresses A-MIPS and enhances swarming. 

\begin{figure}[t]
\centering
  \includegraphics[width= \columnwidth]{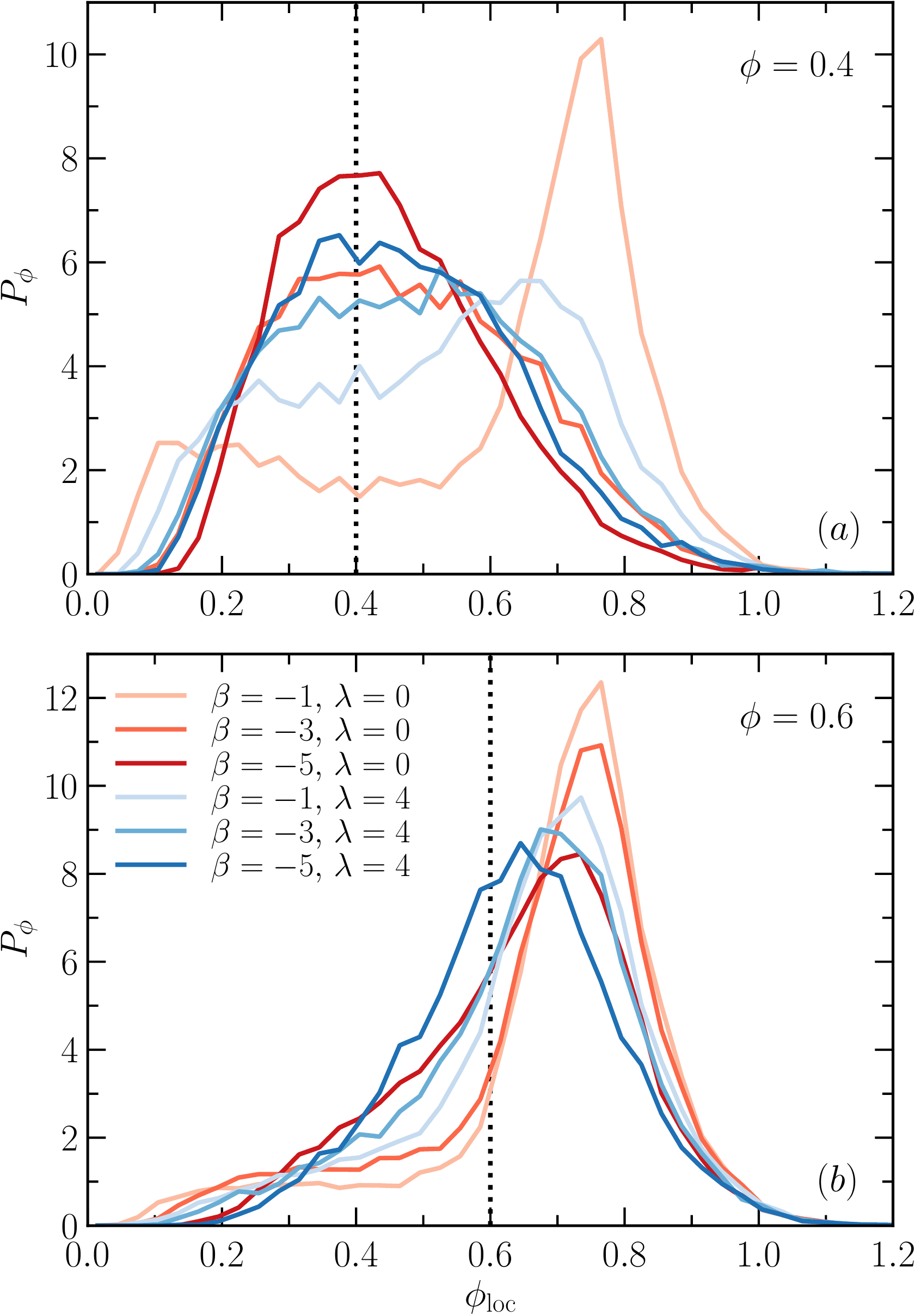} \\
  \caption{{\bf Local packing fraction.} Probability distribution $P_{\phi}$ of the local packing fraction $\phi_{\rm loc}$ for the average area packing fraction (a) $\phi = 0.4$  and (b) $\phi = 0.6$ (vertical dotted lines). The various curves correspond to  $\beta = -1$, $-3$, and $-5$ (bright to dark), and $\lambda = 0 $ (red) and $\lambda = 4$ (blue), respectively.    }
  \label{fig:voroni}
\end{figure}

\subsubsection*{Local packing fraction}

Clustering and A-MIPS  of the squirmers are analyzed  quantitatively  by  a Voronoi tessellation of the accessible volume \cite{rycr:09,wyso:14,thee:18,pers:04}. Figure~\ref{fig:voroni}  provides examples of density distributions for the average packing fractions $\phi =0.4$ and $0.6$. The pronounced peak at the local packing fraction $\phi_{loc} \approx 0.75$ for $\phi =0.4$, $\beta =-1$, and $\lambda=0$ indicates  A-MIPS, with a dense phase in contact with a  dilute phase, consistent with the snapshots of Fig.~\ref{fig:phase_diagram}. Results for large $|\beta|$ imply a disintegration of the large aggregate and ultimately, for $\beta  < -3$,  $P_{\phi}$  displays a maximum at the average packing fraction, which indicates the absence of phase separation.  Similarly, at $\phi =0.6$, the peaks in Fig.~\ref{fig:voroni}(b) for $\lambda =0$ indicate phase separation, even for $\beta$ as negative as $\beta=-5$.    The rotlet dipole prevents formation of large clusters, but even for $\beta =-5$ and $\kappard = 4$ a broad range of cluster sizes exists.

\subsubsection*{Cluster-size distribution } \label{ssec:cluster_size}

The cluster-size distribution function 
\begin{equation} \label{cl_size_distr_eq}
    \mathcal{N}(n) =\frac{1}{N_{sq}}np(n)  
\end{equation}
represents the  fraction of squirmers belonging to a cluster of size $n$, where
$p(n)$ is the number of clusters of size $n$. The distribution is normalized such that 
$\sum^{N_{sq}}_{n=1} \mathcal{N}(n)=1$.  We use a distance and an orientation criterion to define a cluster: a squirmer belongs to a cluster, when its closest distance to another squirmer of the cluster is $d_s < 1.8(2^{1/6}-1) \sigma_s$ and the angle between the orientations of the two squirmers is $< \pi/6$ (see Methods section for the definition). The latter allows us to identify different clusters even at high packing fractions.  

The cluster-size distribution function is a useful quantity to characterize the motility pattern of a microswimmer system \cite{beer:20,levi:14}. In the homogeneous phase, the distribution function decays exponentially,  whereas a second peak (bimodal distribution) indicates the formation of giant clusters (A-MIPS).  At the percolation transition,  $\cal{N}$ becomes scale free and decays by a power law, $\mathcal{N} \sim  x^{-\gamma}$ \cite{levi:14}.  The swarming phase is characterized by a power-law decay with an exponential cut-off and  a characteristic scale determined by an average vortex size \cite{beer:20}.
The distribution functions  presented in Fig.~\ref{fig:cluster_size_distr} confirm our above conclusions on the emergent phases and motility patterns.

For $\phi =0.4$ and $(\beta,\kappard) = (-1,0), (-1,4)$,  $\phi =0.6$, $\kappard=0$, and all considered $\beta$, as well as $(\beta,\kappard) = (-1,4), (-3,4)$, we obtain  bi- and multimodal  distributions with a power-law decay (cf. Tab.~\ref{tab:cluster_size}) at small  cluster sizes 
and a high probability for giant clusters ($N_{sq}=270, \  \phi =0.4$ and $N_{sq}=833, \  \phi =0.6$). This indicates  A-MIPS \cite{thee:18,beer:20}.  The large polar  clusters are mobile, but the systems lack the characteristic large-scale swirling patterns of swarming (cf. movies M1 and M2).  The distribution functions for $\phi =0.4$, $(\beta, \kappard) = (-5, 0), (-5, 4)$ decay in a qualitative different manner. They are well fitted by the function \cite{alar:17}
\begin{align} \label{eq:cluster_size_fit}
\mathcal{N}(x) = A x^{- \gamma}  e^{-x/x_1} .
\end{align} 
This functional form  is observed in various cluster-forming processes \cite{levi:14}. The function interpolates between the power-law decay found for percolating clusters and an exponential suppression of larger clusters.   Table~\ref{tab:cluster_size} presents the fit parameters for the  various curves  of Fig.~\ref{fig:cluster_size_distr}. 
The exponential large-$n$ decay for $\phi =0.4$, $(\beta, \kappard) = (-5, 0), (-5, 4)$ with a  small value of $x_1$ reflects the predominance of  very small clusters --- such  systems are considered as a  gas of clusters.  In contrast, the cluster-size distribution for  $\phimean=0.6$, $(\beta, \kappard) = (-5, 4)$   decreases  over a broad range of $n$ in a power-law fashion reflecting the presence of a wide distribution of cluster sizes ($x_1 \approx 80$), and only larger clusters are exponentially suppressed --- this system is in the swarming phase. The major difference to systems with  $(\beta, \kappard) = (-1, 4), (-3, 4)$ at this concentration is the more pronounced suppression of large clusters, which renders the overall system more dynamic.  


 \setlength{\tabcolsep}{0.5em}
\begin{table} 
\caption{{\bf Fit parameters of cluster size distribution.} Parameters of the cluster-size distribution  function, Eq.~\eqref{eq:cluster_size_fit}, for various average squirmer densities, active stresses, and rotlet dipole strengths. ``$/$''  indicates absence of the exponential function, i.e.,  ${\cal N} =A x^{-\gamma}$. The last column classifies the systems according to their structures and collective behavior. No entry indicates  inconclusive behavior. }
\begin{tabular}{c c c | c c c |c }
$\phimean$ & $\beta$ & $\kappard$ & $A$ & $x_1$ & $\gamma$ & mode  \\ \hline
$0.4$ & $-1$ &  $0$ & 0.17 & $/$ &  2.0  & A-MIPS\\
$0.4$ & $-3$ &  $0$ & 0.86 & $/$ & 1.8  &  \\
$0.4$ & $-5$ &  $0$ & 0.75 & 1.38 & 0.31 &  clus. gas \\
$0.4$ & $-1$ &  $4$ & 0.5 & $/$ & 1.8  &  A-MIPS\\
$0.4$ & $-3$ &  $4$ & 1.25 & $/$  & 2.3  &   \\
$0.4$ & $-5$ &  $4$ & 0.98 & 2.0 & 0.7 &  clus. gas \\ \hline
$0.6$ & $-1$ &  $0$ & 0.08 & $/$ & 1.5 & A-MIPS  \\
$0.6$ & $-3$ &  $0$ & 0.14 & $/$ & 1.5 & A-MIPS  \\
$0.6$ & $-5$ &  $0$ & 0.34 & $/$ & 1.4  & A-MIPS  \\
$0.6$ & $-1$ &  $4$ & 0.22 & $/$ & 1.4  & A-MIPS\\
$0.6$ & $-3$ &  $4$ & 0.27 & $/$ & 1.4   & A-MIPS\\
$0.6$ & $-5$ &  $4$ & 0.32 & 80 & 1.25   & swarming\\
\end{tabular}
\label{tab:cluster_size} 
\end{table} 

The probability distribution functions of the local packing fraction (Fig.~\ref{fig:voroni}) and cluster-size distribution functions (Figs.~\ref{fig:cluster_size_distr}) clearly reveal a marked effect of the rotlet dipole on the collective behavior of the squirmers. In particular, A-MIPS is suppressed, but  formation of highly dynamic  clusters  prevails, with a rather broad distribution of cluster sizes for high squirmer densities.

\subsection*{Dynamical properties} \label{sec:dynamics}

\begin{figure}[t]
\centering
  \includegraphics[width= \columnwidth]{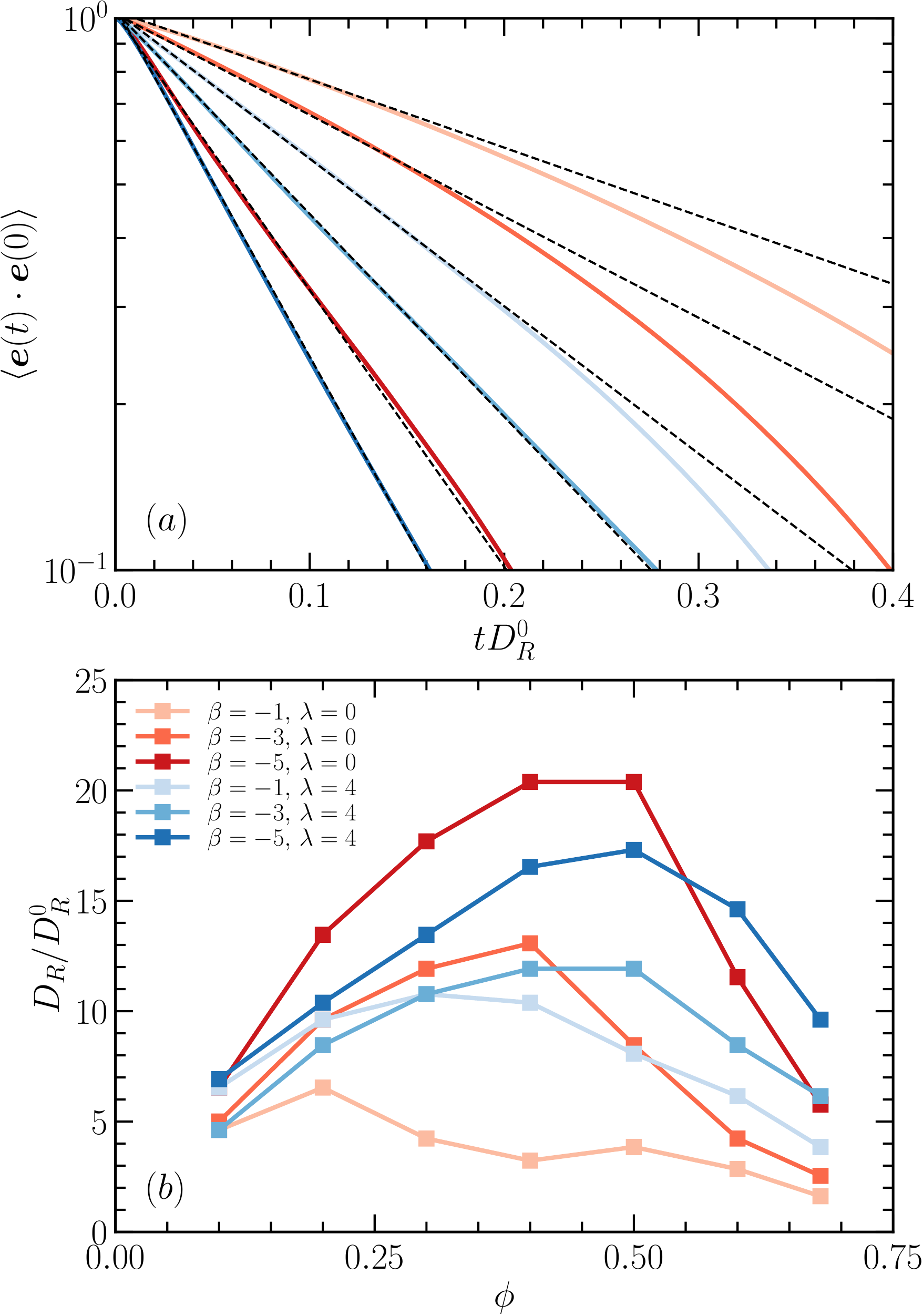} 
  \caption{{\bf Propulsion direction autocorrelation function.} (a) Autocorrelation function of the propulsion  direction as a function of time for the packing fraction $\phimean = 0.6$. $D_R^0$ is the diffusion coefficient  of an individual squirmer in the slit.  The dotted lines are fits to  Eq.~\eqref{eq:diff_coeff_fit}   
  (b) Diffusion coefficients, $D_R$, obtained by a fit of Eq.~\eqref{eq:diff_coeff_fit} as a function of the average packing fraction $\phimean$.  The curves  indicate results for $\beta = -1$, $-3$, and $-5$ (bright to dark) and $\lambda = 0 $ (red) and $\lambda = 4$ (blue), respectively.   }
  \label{fig:rotational_diffusion}
\end{figure}

\subsubsection*{Rotational diffusion}

An individual squirmer in the slit  exhibits rotational diffusion around a minor body axis. Interactions between squirmers, either steric or  by their flow fields, change their diffusive behavior substantially \cite{thee:16.1,thee:18}. Figure~\ref{fig:rotational_diffusion}(a) displays the time dependence of the autocorrelation function $\langle \bm e(t) \cdot \bm e(0) \rangle$ of the propulsion direction of the squirmers. The various curves reflect a marked dependence of the rotational dynamics on the active stress and the rotlet dipole strength.   The correlation function of the systems for  $(\beta, \kappard) = (-1,0), (-3,0), (-1,4)$  exhibit a non-single-exponential decay. Steric interactions between squirmers with a preference to cluster formation as well as between finite-size clusters  lead to a rotation of whole clusters, which implies a faster decay of the rotational  correlation compared to thermal fluctuations alone (cf. movie M4) \cite{gino:18}.

We  characterize the rotational motion by fitting the initial decay of the correlation function with the exponential 
\begin{align} \label{eq:diff_coeff_fit}
 C_R(t) = \lla \bm e(t) \cdot \bm e(0) \rra = C_R^0 e^{-D_Rt} ,
\end{align}
as displayed in Fig.~\ref{fig:rotational_diffusion}(a). The factor $C_R^0 \approx 1.03$ is included to account for a non-exponential decay for very short times.  Squirmers with large active stresses and a rotlet dipole ($(\beta, \kappard) = (-5,0),  (-3,4),  (-5,4)$) exhibit an exponentially decaying correlation function of $C_R$ over more than a order of magnitude. The extracted rotational diffusion coefficients $D_R$ obey $D_R/D_R^0 >1$ (Fig.~\ref{fig:rotational_diffusion}(b)), which reveals an accelerated rotational motion  by  shape-induced steric interactions and hydrodynamic flow fields. Note that $D_R^0$ in a dilute system is independent of $\beta$. The diffusion coefficient $D_R$ increases with increasing squirmer concentration, reaches a packing fraction-dependent maximum and decreases again for larger $\phimean$. An increasing number of squirmer contacts with increasing $\phimean$  ($\phi \lesssim 0.5$) leads to a faster reorientation. However, at larger $\phimean$,  clusters are formed, which move collectively and more persistently, which reduces $D_R$. The larger $D_R$ values for larger $|\beta|$ demonstrate  the substantial contribution of active stress to the reorientation of the squirmers. At smaller $\phimean$ and $\beta < -1$, the presence of a  rotlet dipole  with $\kappard =4$ evidently reduces $D_R$ compared to that for $\kappard =0$, which is  associated with the appearance of  small clusters of side-by-side swimming squirmers (cf. Fig.~\ref{fig:phase_diagram} and movie M4). In contrast, at high packing fractions, a rotlet dipole implies a larger $D_R$ as a consequence of an enhanced orientational motion of smaller clusters, specifically at large $|\beta| = 5 $.

\begin{figure}[t]
\centering
  \includegraphics[width= \columnwidth]{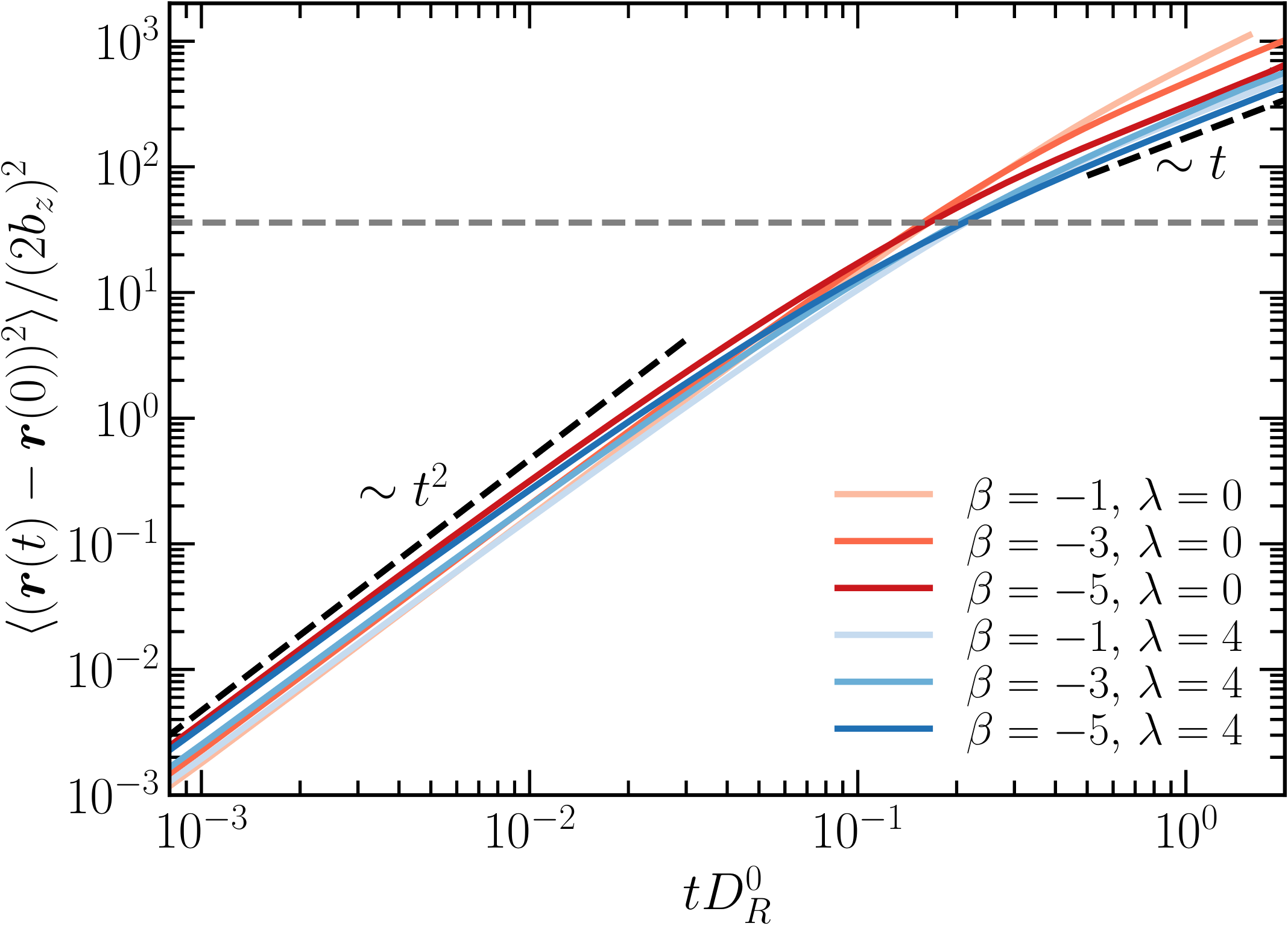} 
  \caption{{\bf Mean-square displacement.} Squirmer mean-square displacement as a function of time for the  packing fraction $\phimean = 0.6$.  The black  dashed lines indicate the power laws of ballistic ($t^2$) and diffusive  ($t$) motion, respectively.  The horizontal gray dashed line corresponds to the displacement of  $6$ squirmer lengths.   }
  \label{fig:mean_square_displ}
\end{figure}

\begin{figure}[t]
\centering
  \includegraphics[width=\columnwidth]{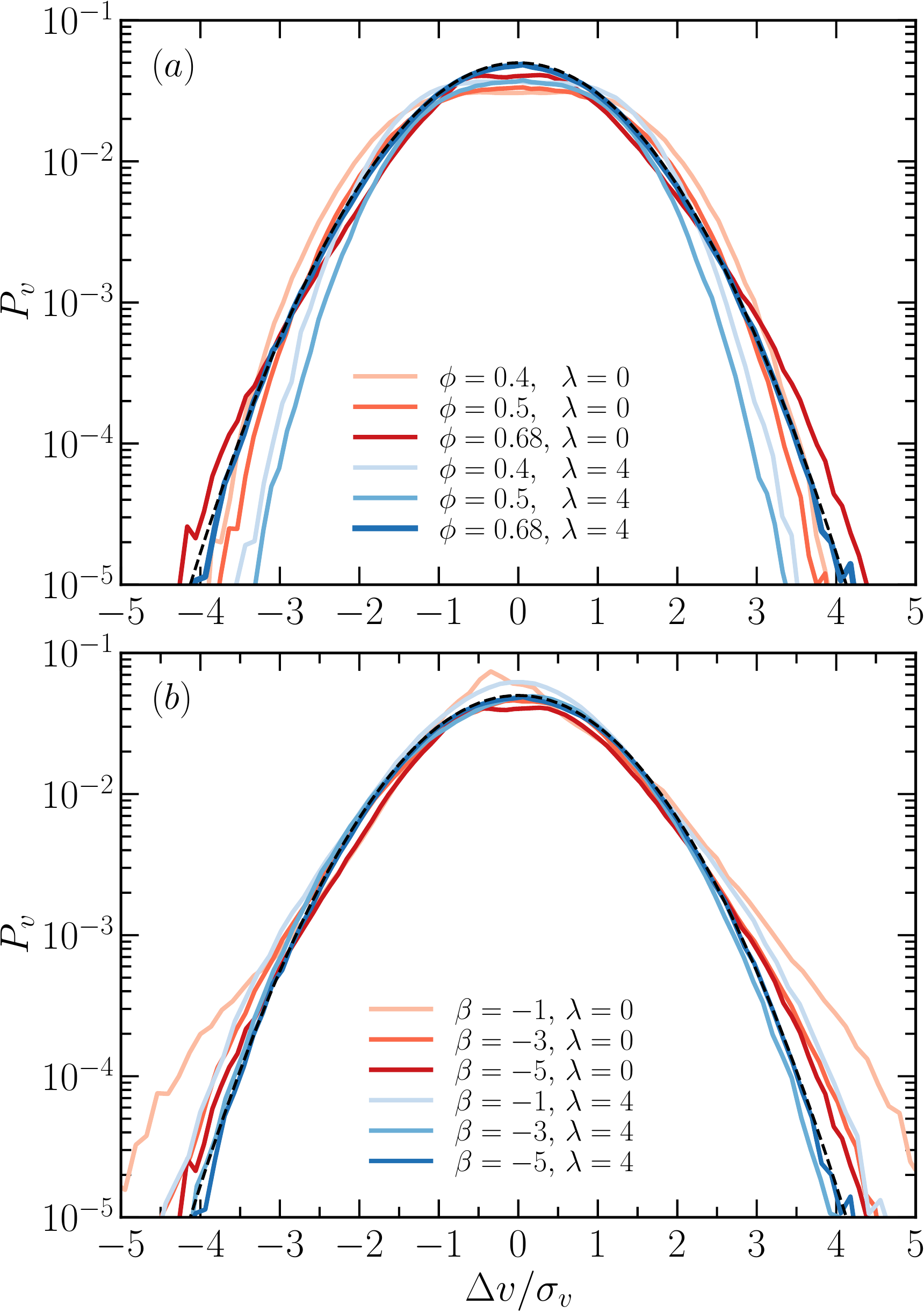} 
  \caption{{\bf Velocity distribution function.} Distribution function $P_v$ of the Cartesian in-plane velocity components ($ \Delta \bm v = (\Delta v_x, \Delta  v_z)^T$) with respect to the mean velocity, normalized by the standard  deviation $\sigma_v$, for (a) $\beta =-5$ and various packing fractions, and (b) $\phimean=0.68$ and various $\beta$ and $\kappard$. The dashed line is the corresponding Gaussian of unit variance.   }
  \label{fig:velocity_distribution}
\end{figure}

\begin{figure*}[t]
  \centering
  \includegraphics[width= \textwidth]{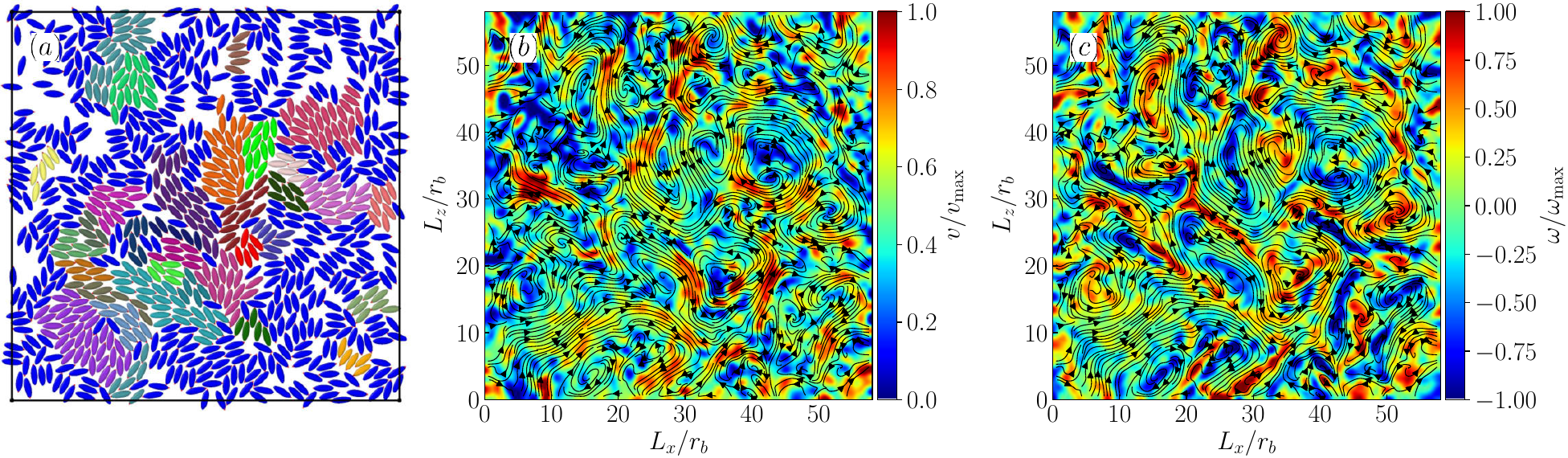}
    \caption{{\bf Squirmer flow fields.} Chaotic collective dynamics of squirmers. (a) Snapshot illustrating the presence of clusters. (b) Velocity field $\bm v (\bm r,t)$  and (c) vorticity field $\omega (\bm r,t) = \partial v_z/\partial x - \partial v_x/\partial z$ of the system with $N_{sq}=833$ squirmers, $\beta=-5$, $\lambda=4$, and the packing fraction $\phi=0.6$. The 
    black lines with arrows indicate the streamlines of the fields. (See movies M3, M4, and M5.)  }
  \label{fig:turbulent_flow_field}
\end{figure*}

\subsubsection*{Mean square displacement}

The mean-square displacement of the squirmers at high packing fractions ($\phimean \geq 0.6$, Fig.~\ref{fig:mean_square_displ})  
exhibit the typical ballistic motion for short times and  a crossover to a diffusive motion for long times $t D_R^0 \gtrsim 0.1$  \cite{elge:15,bech:16}, at least for systems with $\kappard =4$. (The resolution of the long-time behavior of the phase separated systems for $\kappard=0$ requires longer simulations.) There is only a slight difference in the swimming speed of the various squirmers at short times. The presence of a rotlet dipole causes an earlier deviation from a strict ballistic motion  toward a ballistic-like motion with an exponent somewhat smaller than $2$  as time increases compared to squirmers without such a dipole.    
Most remarkable, the systems with $(\beta, \kappard) = (-5,0),  (-1,4), (-3,4),  (-5,4)$ exhibit a crossover from a ballistic or near ballistic to a diffusive motion at a displacement roughly corresponding to $12 b_z$, i.e., $6$ squirmer lengths.  We may consider this as a characteristic length scale in the system, separating the scale of persistent motion from that of diffusive motion. 

\subsubsection*{Velocity distribution function}

Thermal and active fluctuations imply strongly varying instantaneous squirmer velocities, with magnitudes exceeding the swimming velocities by far. Hence, for the calculation of the velocity distribution function, we  determine a swimming velocity by the difference quotient
\begin{align} \label{eq:swim_velocity}
\bm v_i(t) = \frac{\bm r_i(t) -\bm r_i(t-\Delta t)}{\Delta t} .
\end{align}    
During the selected time interval $\Delta t = 10^3\sqrt{ma^2/(k_BT)}$, a squirmer moves at most the distance $2b_z/3$. 

The distribution function $P_v$ of the  Cartesian in-plane velocities $\Delta  v =  (v_{x/z} - {\bar v}_{x/z})$ --- the two spatial dimensions are equivalent --- , where ${\bar v}_{x/z}$ are the average velocities along the Cartesian directions $x$ and $z$, of an single squirmer (dilute system) in the slit, is Gaussian due to the thermal noise of the fluid. (The velocities ${\bar v}_{x/z}$ are typically very small and non-zero only due to finite-size effects and statistical inaccuracy.) Collective effects modify the distribution function and $P_v$ deviates  from a Gaussian in general, as shown in Fig.~\ref{fig:velocity_distribution}. Even for a pronounced active stress, $\beta=-5$, the distribution functions for packing fractions $\phimean < 0.6$  deviate from a Gaussian (Fig.~\ref{fig:velocity_distribution}(a)) independent of $\kappard$ --- the curves are flattened at the maximum and are wider or narrower in the tails. Similar, at $\phimean =0.68$ (Fig.~\ref{fig:velocity_distribution}(b)), $P_v$ is broadened for all systems with $\kappard=0$, as well as for $\beta=-1$ and $\kappard=4$, although the distribution function are close to a Gaussian.  

Remarkably, the velocity distribution function  for the  system with $\phimean=0.68$ and $(\beta, \kappard) = (-5,4)$ is very well described by a Gaussian despite pronounced  collective swimming.   Evidently, steric and flow-field interactions induce sufficient randomness  to yield isotropic  two-dimensional Gaussian distributed velocities. This aspect is particularly relevant for active turbulence, because velocities in high-Reynolds-number turbulent flows are  Gaussian  distributed \cite{wens:12,dunk:13}

\subsection*{Active turbulence} \label{sec:turbulence}

The characteristic features  of the swimmer flow fields at higher densities are illustrated in Fig.~\ref{fig:turbulent_flow_field}. 
The clusters depicted in Fig.~\ref{fig:turbulent_flow_field}(a) exhibit a chaotic  collective   motion with regions of low and high velocity (Fig.~\ref{fig:turbulent_flow_field}(b)) and vorticity  (Fig.~\ref{fig:turbulent_flow_field}(c)) (see movies M3, M5, and M6 for the packing fraction $\phi=0.68$). The patterns are similar to those observed in experiments on bacteria \cite{domb:04,soko:12,dunk:13,wens:12,beer:20}, previous simulations \cite{wens:12,gros:14}, and continuum theory \cite{wolg:08,wens:12,dunk:13}.

\subsubsection*{Spatial velocity correlation function}

\begin{figure}[t]
 \centering
   \includegraphics[width= \columnwidth]{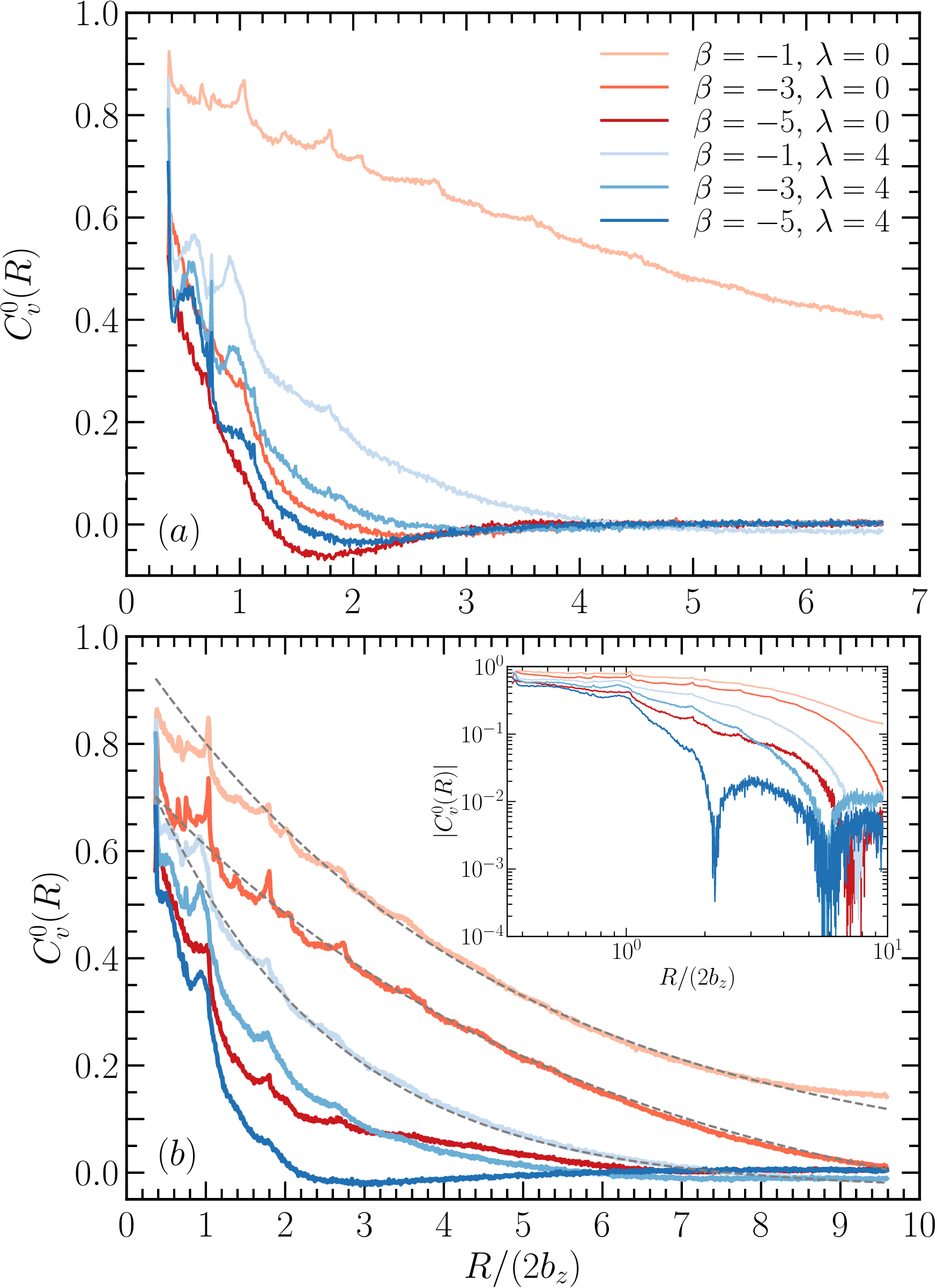}
     \caption{{\bf Velocity correlation function.} Normalized spatial velocity correlation function $C_v^0(R)$  for  the packing fraction (a) $\phimean = 0.4$ and (b) $\phimean = 0.6$ . The inset in (b) displays $C_v^0$ in log-log representation. Dashed lines are fits to Eq.~\eqref{eq:corr_function}. }
   \label{fig:vel_corr_func}
 \end{figure}

Quantitative insight into the turbulent dynamics of the squirmers is obtained by their spatial velocity correlation function, a concept  well established in classic hydrodynamic turbulence \cite{kolm:91,fris:95,batc:59,wens:12}. For the discrete particle system, we define the spatial velocity correlation function as
\cite{wyso:14,wens:12.1,chen:12} 
\begin{align} \label{eq:corr_function}  
C_v(\bm R)=\frac{\left\langle \sum_{i,j\neq i} \bm v_i (t) \cdot   \bm v_j (t)  \delta(\bm R-|{\bm r}_i-{\bm r}_j|)\right\rangle}{\left \langle \sum_{i,j\neq i} \delta(\bm R-|{\bm r}_i - {\bm r}_j|) \right \rangle} ,
\end{align}
where $\bm r_i$ is the center-of-mass position of squirmer $i$. 
Moreover, we introduce a normalized velocity correlation function as $C_v^0(R) = C_v(R)/c_0$,
with $c_0 = \sum_i \langle \bm v_i^2 \rangle/\nsq$. (For an homogeneous and isotropic system, $C_v(\bm R)$ is a function of $R =|\bm R|$ only.) Results of $C_v^0$  for the  packing fractions $\phimean =0.4$ and $0.6$ are presented in Fig.~\ref{fig:vel_corr_func}. Three distinct decay patterns can be identified: (i) a very slow decay over roughly the whole system  ($\phimean=0. 4$, $(\beta , \kappard) = (-1,0)$; $\phimean=0. 6$, $(\beta, \kappard) =(-1, 0), (-3,0)$), (ii) a decay, where  correlations functions are negative for   $R \lesssim  L/2$ ($\phimean=0. 4$, $(\beta , \kappard) = (-1,4) $; $\phimean=0. 6$, $(\beta , \kappard) = (-1,4), (-3,4)$), and (iii) correlations functions, which assume negative values over a certain interval, but are positive for $R\approx L/2$ ($\phimean=0. 4$, $(\beta , \kappard) = (-5|0), (-3,4), (-5,4) $; $\phimean=0. 6$, $(\beta , \kappard) = (-5,4)$).  The case (i) corresponds to  long-range correlations over the entire simulation box, consistent with A-MIPS and the appearance of a large cluster (Fig.~\ref{fig:cluster_size_distr}).  As shown in Fig.~\ref{fig:vel_corr_func}(b), such $C_v^0(R)$ can be fitted by the function 
\begin{align} \label{eq:corr_vel_fit}
 C_v^0(x) = A_v  e^{ - x/\xi  }  - g . 
\end{align}
Specifically for $\phimean=0.6$, we obtain the parameters of Tab.~\ref{tab:vel_corr}.  The respective velocity correlation functions  decay approximately exponentially, with  characteristic lengths scales between  $2.3$ and $5.6$ swimmer lengths.  The smaller value $\xi/(2b_z) = 2.3$ for $\lambda =4$  indicates that a non-zero rotlet dipole implies weaker spatial correlation and, hence, smaller clusters.  
The distinct decay patterns support our conclusion  on the motility mode as discussed in relation the cluster-size distribution functions (Fig. \ref{fig:cluster_size_distr}). However, a clear-cut separation of  swarming and cluster dynamics  is difficult to establish based on $C_v(R)$.

An important feature of bacterial turbulence is a finite vortex size, which marks a characteristic length scale in the system and is reflected in a minimum of the velocity correlation function \cite{zhan:09.1,wens:12,dunk:13,chen:12}. Our simulations yield such a minimum, e.g., for $\phimean=0. 6,  \, 0.68$, $(\beta, \kappard) = (-5,4)$. Hence, we expect such squirmer system  to exhibit active turbulence.  A characteristic length scale can also exist for lower densities, e.g., for  $\phimean=0. 4$, $(\beta, \kappard) = (-5,0) , (-5,4)$, where only small clusters are present. We would not denote the dynamics of such systems as turbulent.  
 
 \setlength{\tabcolsep}{0.5em}
 \begin{table} 
 \caption{{\bf Fit parameters of velocity correlation function.} Parameters of the spatial velocity correlation  function, Eq.~\eqref{eq:corr_vel_fit}, for various active stresses and rotlet dipole strengths, and the squirmer density $\phi=0.6$.}
 \begin{tabular}{c c |c  c c  }
  $\beta$ & $\kappard$ & $A_v $ & $\xi/(2b_x)$  &  $g$  \\ \hline
  $-1$ &  $0$ & 1.00 & 4.5 &  0.00\\
  $-3$ &  $0$ & 0.92 & 5.6 &   0.16 \\
  $-1$ &  $4$ & 0.86 & 2.3 &  0.032\\
 \end{tabular}
 \label{tab:vel_corr} 
 \end{table} 
 
\begin{figure}[t]
\centering
  \includegraphics[width= \columnwidth]{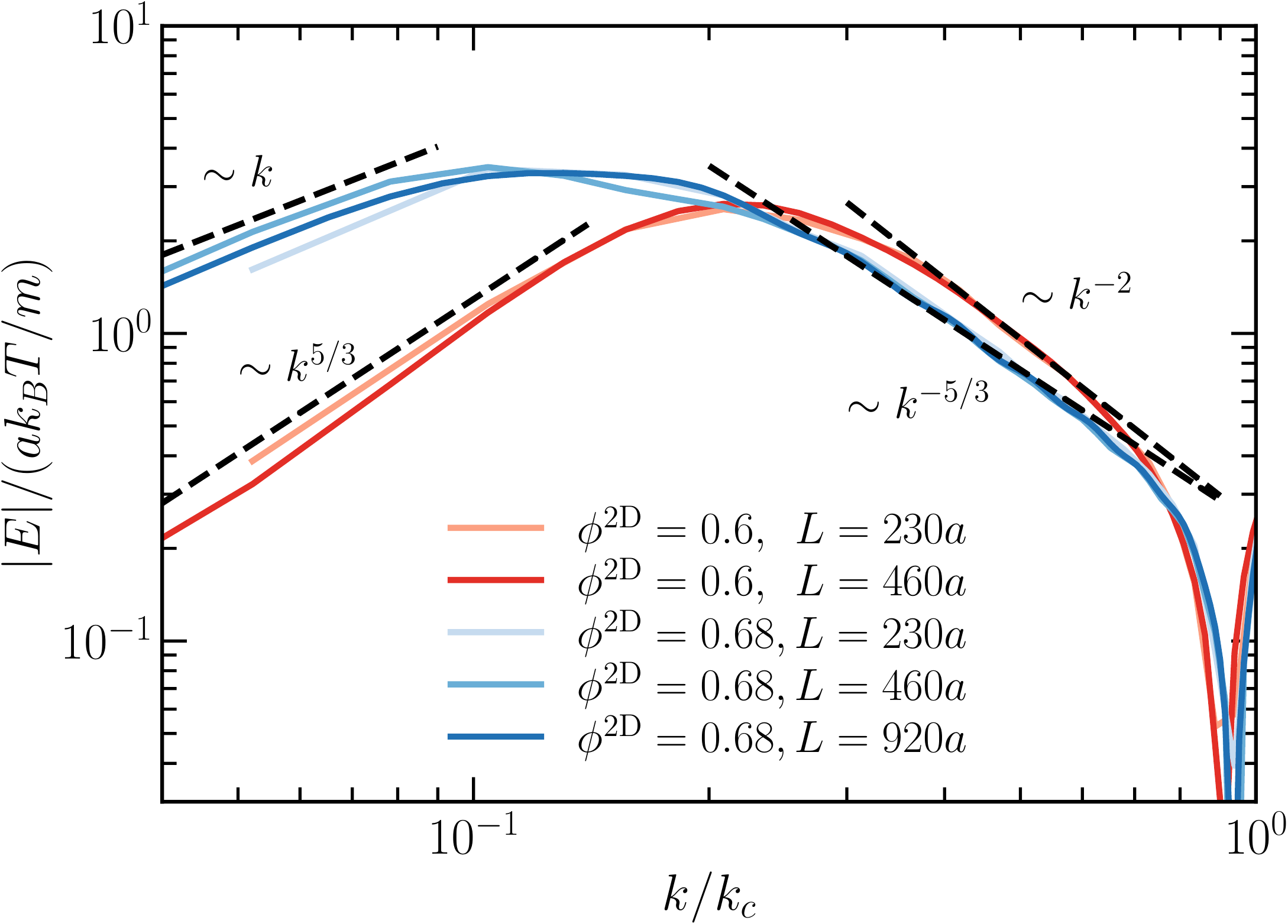}
    \caption{{\bf Energy spectrum.} Energy spectra of systems with $\beta=-5$ and $\kappard=4$ for the packing fractions 
        $\phimean = 0.6$ (red) and $0.68$ (blue). Various system sizes (see legend) have been explored in order to verify absence of finite-size effects. The dashed lines indicate power-laws in the respective regimes.  }
  \label{fig:energy_spectrum}
\end{figure}

\subsubsection*{Energy spectrum}

Insight into the turbulent behavior is gained by the energy spectrum 
\begin{align} \label{eq:eng_spectrum}
    E(k)=\frac{k}{2\pi}\int d^2 R \; e^{-i {\bm k} \cdot {\bm R}} C_v(\bm R) , 
\end{align} 
which is obtained as Fourier transform of the spatial velocity correlation function  \eqref{eq:corr_function} \cite{batc:59}, and manifests the distribution of kinetic energy over different length scales. In the calculation of $E(k)$, we apply
a left-shift of the correlation function $C_v(R)$  (Fig.~\ref{fig:vel_corr_func}) such that  the decay starts at $R=0$  in order to avoid artifacts in the Fourier transformation by a truncated correlation function. As for bacterial suspensions, the  energy injection scale is the length scale of a microswimmer ($2b_z$), which yields the characteristic (maximum) wavenumber $k_c=\pi/b_z$ for our squirmers.

Figure~\ref{fig:energy_spectrum} displays the energy spectrum for $(\beta, \kappard)  = (-5, 4)$ and the two  packing fractions $\phimean =0.6$, $0.68$, and various system sizes. The  simulations show two power-law  regimes for a given density, namely  $E(k) \sim k^{\hat \kappa}$ for $k< k_{\rm m}$ and $E(k) \sim k^{- \kappa}$  for $k_{\rm m}< k< k_c$, with $k_{\rm m}$ corresponding to the peak position of $E(k)$. Such a maximum in $E(k)$ is a feature of  microswimmer active turbulence, and reflects a characteristic vortex size \cite{wens:12,soko:12,dunk:13}. Our simulations yield  approximate vortex sizes of $5$  ($10 b_z$)  and $10$ squirmer lengths ($20 b_z$) for $\phimean = 0.6$ and $0.68$, respectively. They are roughly consistent with the  patterns of Fig.~\ref{fig:turbulent_flow_field}, the crossover from ballistic to diffusive motion in the mean-square displacement of Fig.~\ref{fig:mean_square_displ}, and the minimum of the correlation function of Fig.~\ref{fig:vel_corr_func}(b).  Vortex sizes on the order of $5 -10$ microswimmer lengths are also found in  experiments \cite{wens:12,soko:12,bepp:17}. 

For $k_{\rm m}< k< k_c$, corresponding to $R > 2b_x$, our simulations yield turbulent flow patterns (Fig.~\ref{fig:turbulent_flow_field}). The exponent of the scaling regime depends on the squirmer density, with the values $\kappa = -2$  for $\phimean=0.6$ and $\kappa = -5/3$ for $\phimean =0.68$. The latter is consistent with the Kolmogorov-Kraichnan  prediction for classical 2D turbulence \cite{krai:80}. This is remarkable, considering the wide scatter of exponents found in simulations and  experiments  (cf. Tab.~\ref{tab:energy_spectrum}). Density seems to play an important role for the observed  turbulent behavior. The squirmers of both densities exhibit swarming, namely, collective motion with large-scale swirling and streaming patterns.  However, only the dynamics in the higher density system exhibits the exponent $\kappa=5/3$.  

In the small $k$-value regime, we obtain the exponents  $ \hat \kappa = 1$ for $\phimean=0.68$ and $ \hat \kappa = 5/3$ for $\phimean=0.6$,  which reflect  an increase of the energy with increasing $k$.  The dependence $k^{5/3}$  is consistent with that observed  theoretically and experimentally in Ref.~\cite{wens:12}, as well as in
simulations \cite{gros:14}. However, other studies yield rather different dependencies (Tab.~\ref{tab:energy_spectrum}). Theoretical models suggest that the small-$k$ slope is governed by  finite-system-size effects, i.e., depends in the boundary condition and physical parameters \cite{brat:15}.  The curves in Fig.~\ref{fig:energy_spectrum} reflect  a weak dependence on the system size. 

The presences  of a small-distance cut-off, where energy input by the squirmers occurs, and the  peak in $E(k)$, corresponding to a characteristic vortex size, limits the $k$-range over which the energy spectrum decays in a power-law manner. This is in stark contrast to classical high-Reynolds-number turbulence, where the energy cascade extents over many orders of magnitude.

\section*{Conclusions} \label{sec:summary}

We have performed large-scale mesoscale hydrodynamics simulations of spheroidal squirmers in a narrow slit in order to analyze the emerging  structures, motility patterns, and turbulent behavior for various packing fractions, active stresses, and rotlet-dipole strengths. 

Our studies reveal a strong dependence of the motility pattern on the microswimmer concentration and their propulsion-induced flow field. The classification of the distinct motion pattern into the various categories --- swimming and collective motion of very small clusters (cluster gas),  phase separation by activity and anisotropic swimmer shape  (A-MIPS), and swarming --- is  accomplished by visual inspection of snapshots (Fig.~\ref{fig:phase_diagram}) and the characteristic features of the  cluster-size distribution function (Fig.~\ref{fig:cluster_size_distr}).  A-MIPS appears for small active stresses, $|\beta| \lesssim 3$, and all packing fractions  $\phimean >0.2$. Squirmers with  stronger forces dipoles, $|\beta| \gtrsim 3$, at concentrations  $\phimean < 0.4$ exhibit small clusters and strong cooperative effects for $\kappard =4$. At higher packing fractions, $\phimean \geq 0.5$, a swarming phase appears for  $\kappard =4$, where clusters of squirmers move collectively, and even exhibit  active turbulence for high packing fractions  ($\phi=0.6, 0.68$) and sufficiently large $|\beta|$ (Fig.~\ref{fig:phase_diagram}). Importantly, the rotlet dipole suppresses A-MIPS.   

Our simulations clearly reveal the difficulty to characterize turbulence in active systems. Even more fundamental  is the question, which criteria should be applied to classify a mesoscale system as  turbulent,  Considering microswimmer systems, chaotic flow patterns are evidently not sufficient. Inspired by classical hydrodynamic turbulence, we propose the  following  ``minimal'' criteria:
\begin{itemize}
\item Reynolds numbers $Re < 1$
\item presence of chaotic flow patterns with large-scale collective behavior 
\item characteristic vortex size and a negative velocity correlation function
\item Gaussian velocity distribution function of the microswimmer's Cartesian velocity components
\item energy spectrum with power-law decay $E(k) \sim k^{-\kappa}$, $\kappa >0$, on length scales below the characteristic vortex size.
\end{itemize}
The presence of small and large length-scale  cut-offs by the microswimmer and vortex size  implies a universal, scale-free behavior only over a limited range of  length scales. 

Analyzing the swarming motion of the squirmers, we find non-Gaussian distribution functions for the velocities parallel to the confining walls for $\phimean < 0.6$. According to our criteria, we classify such systems as non-turbulent. Yet, we obtain a Gaussian velocity distribution for $\phimean=0.68$ and $(\beta, \kappard) = (-5, 4)$ (Fig.~\ref{fig:velocity_distribution}). The energy spectrum of that system exhibits a power-law decay with the exponent $\kappa = 5/3$, characteristic for Kolmogorov-Kraichnan-type turbulence in the inertial range. Hence, this systems fulfills all the above criteria, and we consider it as fully turbulent.  

The slope of the power-law regime depends on  the squirmer density. At the smaller packing fraction $\phimean  =0.6$ and $(\beta, \lambda) = (-5,4)$, the energy spectrum decreases faster, with the exponent $\kappa = 2$. At the same time, the velocity distribution function is non-Gaussian.  Hence, the  system is not showing active turbulence in the above sense, yet, exhibiting swarming motility.  This suggests a tight link between the energy spectrum and the velocity distribution function, a relation which needs further considerations.

As typically observed in turbulent bacterial suspensions \cite{soko:12,dunk:13,wens:12}, we also obtain a maximum in the energy spectrum at $5-10$ squirmer lengths, as well as a negative spatial velocity correlation function, in agreement with the presence of a characteristic vortex size.  

Inertia of the collective active motion could play an important role, since the crossover from the active ballistic motion --- equivalent to inertia of a passive system ---  to active diffusion appears on the length scale of approximately $6$ squirmers lengths, which is comparable to the characteristic vortex size.  Yet, the Reynolds number on the scale of a vortex (approximately $10$ microswimmer lengths) is  still smaller than unity. Here, more detailed theoretical studies of a suitable model are required to assess the relevance of the various interactions on active turbulence.

Despite the similarities of our squirmer systems with bacterial suspensions, there is one major difference, namely, the swimming speed of bacteria increasing in the swarming phase, whereas it decreases in our case \cite{swie:13}. This may point toward a particular role of bacterial flagella in the propulsion of the dense bacterial system. 

We like to emphasize  that hydrodynamic interactions are  paramount for microswimmer swarming and active turbulence, specifically the active stress and the rotlet dipole determine their swarming behavior.  However, for Kolmogorov-Kraichnan-type characteristics to merge, in addition, density plays a major role, and ensures an isotropic and homogeneous dynamics on lengths scales larger than approximately a squirmer length.  Our simulations provide a benchmark for further theoretical and simulation studies on bacterial turbulence to elucidate  the interplay between hydrodynamic stress --- specifically a rotlet dipole ---, alignment interactions by anisotropic swimmer shapes, and volume exclusion.  
 

\section*{Method} \label{sec:method}

\subsection*{Microswimmer model: prolate squirmer} 

The  prescribed surface velocity of the  prolate spheroidal squirmer,  a homogeneous colloidal particle of mass $M$, is given by  the  \cite{ishi:06,pago:13,thee:16.1,zoet:18}
\begin{align} \label{surf_vel}
    {\bm u}_{\rm s} = - B_1 (1+ \beta \zeta)({\bm e}_{\zeta}\cdot {\bm e}){\bm e}_{\zeta}+ \frac{3 \lambda z_s\bar{r}_s}{r_s^5} {\bm e}_{\varphi}
\end{align}
in terms of spheroidal coordinates $\tau, \zeta, \varphi$ ($1 \le \tau < \infty$, $-1\le \zeta \le 1$, $0 \le \varphi < 2 \pi$) (Fig.~\ref{fig:sketch}(a)) \cite{thee:16.1,qi:20}. For a squirmer with propulsion direction $\bm e  = (0,0,1)$, the Cartesian coordinates of a point on the spheroid surface $\bm r_s = (x_s,y_s,z_s)^T$ are 
\begin{align} \nonumber
x_s = & \ b_x \sqrt{1- \zeta^2} \cos \varphi  ,  \ 
y_s = & \ b_x \sqrt{1- \zeta^2} \sin \varphi ,  \ 
z_s = & \ b_z \zeta ,
\end{align}   
with ${\bar r}_s = \sqrt{x_s^2+y_s^2}$, $r_s = |\bm r_s|$, and $\tau = \tau_0=b_z/\sqrt{b_z^2-b_x^2}$ and the lengths $b_z$ and $b_x$ along the  semi-major and -minor axis (Fig.~\ref{fig:sketch}(a)). The terms with the coefficients $B_1$ and $\beta$ ($\beta < 0$, pusher) account for swimming in the direction $\bm e$ and an active stress, respectively \cite{thee:16,thee:16.1,qi:20}. The rotlet-dipole term of strength $\lambda$ accounts for the torque-free nature of swimming bacteria with a counter rotating cell body compared to the rotating flagellar bundle \cite{hu:15.1}. The swimming  velocity of  a squirmer is related to $B_1$ as 
\begin{align} \label{eq:swimm_velo}
    v_0=B_1\tau_0 [\tau_0-(\tau_0^2-1){\rm coth}^{-1}\tau_0] .
\end{align}

To insure  quasi-two-dimensional motion between the walls (Fig.~\ref{fig:sketch}), a strong repulsive interaction between squirmers and walls is implemented by the truncated and shifted  Lennard-Jones potential 
\begin{equation} \label{eq:repl_wall}
    U_w = 4\epsilon_w\left[  \left( \frac{\sigma_w}{y} \right)^{12} - \left( \frac{\sigma_w}{y} \right)^6  + \frac{1}{4} \right] 
\end{equation}
for  $y < 2^{1/6}\sigma_w$ and zero else, where $y$ is the closest distance between a wall  and the  surface of a squirmer. Here, $\sigma_w$ and $\epsilon_w$ determine to the length and energy scale, respectively. 

Squirmer volume-exclusion interactions are described  by  a separation-shifted  Lennard-Jones potential   with parameters $\sigma_s$ and $\epsilon_s$, where $y \rightarrow d_s + \sigma_s$ in Eq.~\eqref{eq:repl_wall}, and $d_s$ is the distance between the  two closest points on the surfaces of two interacting spheroids \cite{thee:16.1,qi:20}. 

The solid-body equations of motion of the squirmers --- the center-of-mass translational motion and the rotational motion described by quaternions --- are solved by the velocity-Verlet algorithm \cite{thee:16.1,qi:20}.  

\subsection*{Fluid model: Multiparticle collision dynamics} \label{sec:mpc}

The fluid is modeled via the multiparticle collision dynamics (MPC) method, a particle-based  mesoscale simulation approach accounting for thermal fluctuations \cite{kapr:08,gomp:09}, which has  been shown to correctly capture hydrodynamic interactions \cite{huan:12}, specifically for active agents and systems  \cite{gold:09,reig:12,geye:13,brum:14,pak:14,thee:14,eise:16,hu:15,hu:15.1,mous:20,babu:12,rode:19}.

We apply the  MPC  approach with angular momentum  conservation (MPC-SRD+a)  \cite{thee:16,nogu:08}.  The algorithm proceeds in two steps --- streaming and  collision. In the streaming step, the  MPC point particles of mass $m$ propagate ballistically over a time interval $h$, denoted as collision time. In the collision step, fluid particles are sorted into the cells of a cubic  lattice of lattice constant $a$ defining the collision environment, 
and their relative velocities, with respect to the center-of-mass velocity of the collision cell, are rotated around a randomly oriented axes by a fixed angle $\alpha$. The algorithm conserves mass, linear, and angular momentum on the collision-cell level, which implies hydrodynamics on large length and long time scales \cite{kapr:08,huan:12}. A random shift of the collision cell lattice is applied at every collision step to ensure Galilean invariance \cite{ihle:03}. 
Thermal fluctuations are intrinsic to the MPC method.  A cell-level canonical thermostat (Maxwell-Boltzmann scaling (MBS) thermostat) is applied after every collision step, which maintains the temperature at the desired value \cite{huan:10.1}. The MPC method is highly parallel and is efficiently implemented on a graphics processing unit (GPU) for a high-performance gain \cite{west:14}.  

Squirmer-fluid interactions appear during streaming and collision. While streaming  squirmers and fluid particles, fluid particles are reflected at a squirmer's surface by application of  the bounce-back rule and addition of the surface velocity $\bm u_s$ \eqref{surf_vel}.  To minimize slip, phantom particles are added inside of the squirmers, which contribute when  collision cells penetrate squirmers. In all cases, the total linear and angular momenta are included in the squirmer dynamics.
More details are described in Ref.~\cite{thee:16.1} and the supplementary material of Ref.~\cite{qi:20}.

\subsection*{Parameters}

Multiple squirmers with the semi-major axis $b_z=6a$  and semi-minor axis $b_x=2a$  are distributed in a narrow slit of width $L_y=8a$, where $a$ is the length of the MPC fluid collision cell.  Parallel to the walls, periodic boundary conditions are applied.  We set $\sigma_w=1.8a$ and $\epsilon_w=18k_BT$. Squirmer propulsion requires fluid particles adjacent to its surface. To avoid MPC particle depletion when two squirmers approach each other, we introduce a safety layer of thickness $d_v=0.25 a$ around every squirmer, corresponding to the effective squirmer semi-axes $b_z+d_v$ and $b_x+d_v$, respectively. The squirmer-squirmer Lennard-Jones parameters are set to $\sigma_s=0.5 a$, $\epsilon_s=5 k_BT$.  $d_s$ (see microswimmer model) is now the distance between two closest points on the surfaces of the two interacting squirmers with  effective (larger) semi-axes \cite{thee:16.1,thee:18,qi:20}.  

To avoid MPC-particle depletion \cite{thee:18},  we employ a high  average particle number  $\langle N_c \rangle = 60$ in a collision cell. Furthermore, we choose a small collision-time step  $h=0.02 \sqrt{ma^2/(k_BT)}$ and the large rotation angle $\alpha = 130^{\circ}$. This results in the fluid viscosity $\eta = 127.8 \sqrt{mk_BT/a^4}$ and the 2D rotational diffusion coefficient around a minor axis $D_R^0= 5.2 \times 10^{-6} \sqrt{k_BT a^2/m}$. This is in close agreement with the theoretical value of a spheroid  $D_R^0= 5.5 \times 10^{-6} \sqrt{k_BT a^2/m}$.

For  a squirmer, we choose $B_1=0.0045\sqrt{k_BT/m}$, corresponding to the swimming speed $v_0=0.004 \sqrt{k_BT/m}$, which yields the P\'eclet number $Pe = v_0/(2b_z D_R^0)= 64$ and the Reynolds number $Re=2b_zv_0 \langle N_c \rangle/(a^3 \eta) = 0.023$. The  active stress values  $\beta = -1, \  -3, \ -5$, covering approximately the estimated values from experiments and simulations (see below),  and the rotlet dipole strengths $\lambda = 0, \ 4$ are considered.  Simulations with the box size $L =160a$ are performed for the 2D packing fractions $\phi=N_{sq} \pi b_x b_z/L^2 =0.1, 0.2, 0.3, 0.4,$ and $0.5$, corresponding to the squirmer numbers $N_{sq} = 66,  140,  200,  270, $ and $341$. In order to reduce/avoid finite-size effects for higher densities, larger systems are simulated with $L=230 a$ for $N_{sq} = 833, \ 954$,  $L =460$ and $N_{sq} = 3332, \ 3816$ (both  $\phi = 0.6, \ 0.68$), as well as $L = 920$ for  $N_{sq} = 15264$ ($\phi =  0.68$). A passive spheroid  is neutrally bouyant with $M=6031 m$, and the MPC time step $h$  is used in the integration of the squirmers' equations of motion.

\subsection *{Estimation of squirmer parameters for {\em E. coli} from simulations and experiments}

In the far-field, the microswimmer flow field is dominated by the force-dipole term of strength \cite{laug:09,dres:11,elge:15,shae:20}
\begin{align}
\chi = \frac{P}{8 \pi \eta} ,
\end{align}
where $P=f_D l_D$ is the magnitude of the force dipole of force  $f_D$ and length $l_D$. The latter parameters can be determined from experiments \cite{dres:11} and simulations \cite{hu:15.1}. The far-field expansion of the  flow field  of a spheroidal squirmer provides the relation between $\chi$ and the active stress parameter $\beta$ \cite{thee:16.1}:
\begin{align} \label{eq:beta_fd}
\beta = & - \frac{\chi}{v_0 (b_z^2-b_x^2)} \\ \nonumber  & \times \frac{[3 \tau_0 + (1-3 \tau_0^2)\coth^{-1}\tau_0][\tau_0-(\tau_0^2-1)\coth^{-1}\tau_0 ]  }{2/3 - \tau_0^2 + \tau_0 (\tau_0^2-1) \coth^{-1}\tau_0} .
\end{align}
With the approximation of  the bacteria cell body by a spheroid, Eq.~\eqref{eq:beta_fd} provides an estimation of $\beta$ for a given $\chi$.
\begin{itemize}[leftmargin=*]
\item {\em From simulations} --- An {\em E. coli}-type cell model with the  body length $l_b= 2.4 \mu m$, cell body diameter $d_b=0.9 \mu m$, the swimming speed $v_0=40 \mu m/s$,  force-dipole strength $f_D=0.57 pN$, and force-dipole length $l_D= 3.84 \mu m$ \cite{hu:15.1}, yields   $\beta \approx - 6$. 
\item {\em From experiments} --- {\em E. coli} bacteria are characterized by  $l_b=3 \mu m$,  $d_b=1 \mu m$,  $v_0=22 \mu m/s$,  $f_D=0.42 pN$, and  $l_D= 1.9 \mu m$ \cite{dres:11}, which gives $\beta \approx -3 $.
\end{itemize}
In  both cases, the viscosity of water is used. These $\beta$ values approximately fall into the range of  active stresses considered in our simulations.

\section*{Acknowledgments}

This work has been supported by the DFG priority program SPP 1726 ``Microswimmers -- from Single 
Particle Motion to Collective Behaviour''. The authors gratefully acknowledge the computing time 
granted through JARA-HPC on the supercomputer JURECA at Forschungszentrum J\"ulich.

\section*{Data availability}

The data that support the findings of this study are available from the corresponding author upon reasonable request.

\section*{Author contributions}

R.G.W. and G.G. designed the study. K.Q. and E.W. wrote the simulation code and K.Q. performed the simulations. K.Q., R.G.W, and G.G.. analyzed and discussed the results. R.G.W, G.G., and K.Q. wrote the paper.

\section*{Competing interests}

The authors declare no competing interests.

\section*{Additional information}

Supplementary information is available for this paper at ????.


%

\end{document}